# TOWARDS UNDERSTANDING GAMMA-RAY BURSTS

Tsvi Piran
Racah Institute for Physics, The Hebrew University,
Jerusalem, Israel 91904

July 30, 1995

# ABSTRACT

γ-ray bursts (GRBs) have puzzled astronomers since their accidental discovery in the sixties. The BATSE detector on COMPTON-GRO satellite has been detecting GRBs for the last four years at a rate of one burst per day. Its findings has revolutionized our ideas about the nature of these objects. In this lecture I show that the simplest, most conventional and practically inevitable, interpretation of the observations is that GRBs form during the conversion of the kinetic energy of ultra-relativistic particles to radiation. The inner "engine" that accelerates these particles is well hidden from direct observations and its origin might remain mysterious for a long time.

## 0.1 INTRODUCTION

γ-ray bursts (GRBs) were discovered accidentally in the sixties by the Vela satellites. The satellites' mission was to monitor the "outer space treaty" that forbade nuclear explosions in space. A wonderful by-product of this effort was the discovery of GRBs. Had it Not been needed to monitor this treaty, it is most likely that today we would still be unaware of the existence of these mysterious bursts. The discovery of GRBs was announced in 1973 [1]. Since then several dedicated satellites were send to observe the bursts and numerous theories were put forwards to explain their origin. Recently, the BATSE detector of COMPTON-GRO have revolutionized GRB observations and consequently some of our basic ideas on the origin of GRBs. However, BATSE's observations has raised as many new open questions as those they have answered. Some have even said that these observations require "new physics". I examine these questions and directions for their resolution in this lecture.

## 0.2 OBSERVATIONS

GRBs are short non-thermal bursts of low energy γ-rays. The bursts' duration ranges from a few milliseconds to hundred of seconds and the temporal profiles display complicated patterns. After more than twenty years of GRB observations it is still difficult to summarize their basic features. This difficulty stems from the enormous variability displayed by the bursts. I will review here some features, that I believe hold the key to this enigma. I refer the reader to the proceedings of the second Huntsville GRB meeting [2] and to other recent observational reviews for a more detailed discussion [3, 4, 5].

**Duration:** The burst duration ranges from several microseconds to several hundred seconds, with complicated and irregular temporal structure. Several time profiles, selected from the second BATSE 2 catalogue are shown in Fig. 1. The bursts duration distribution is bimodal [6, 7, 8] and can be divided to two sub-groups according to $T_{90}$, the time in which 90% of the burst's energy is observed: Long bursts with $T_{90} > 2$sec and short burst with $T_{90} < 2$sec. Some bursts are extremely long and in one case high energy (GeV) photons have been observed several hours after the main pulse [9]. About 3% of the bursts are preceded by a precursor with a lower peak intensity than the main burst [10].

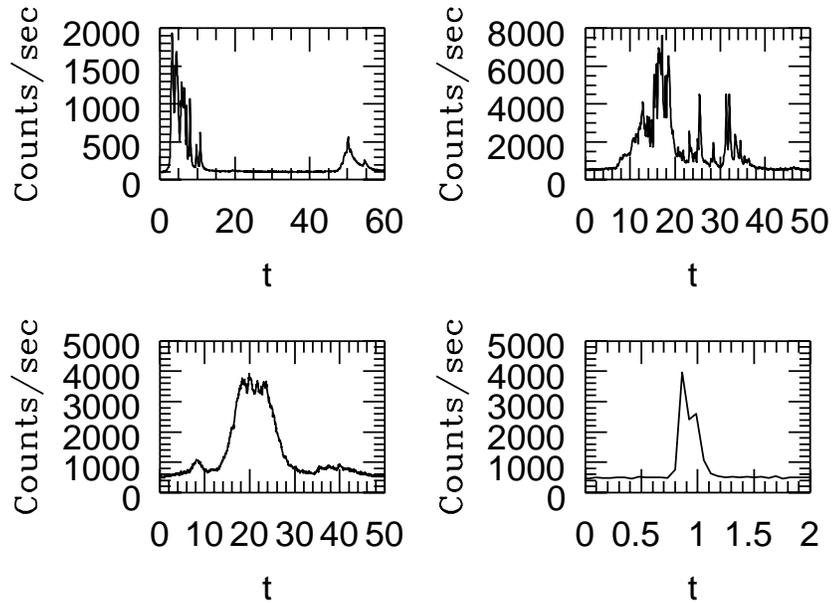

*Fig. 1: Temporal distribution of four bursts from the BATSE 2 Catalogue*

**Spectrum:** Most of the energy of the burst emerges in the several hundred KeV range (see [11] for a recent review). The spectrum is non thermal. The simplest fit is a power law:

$$N(E)dE \propto E^{-\alpha}, \tag{0.1}$$

with a spectral index, $\alpha \approx 2$ (see Fig. 2). Several bursts display high energy tails up to the GeV region. So far BATSE has not found any of the spectral features (absorption or emission lines) reported by earlier satellites [12].

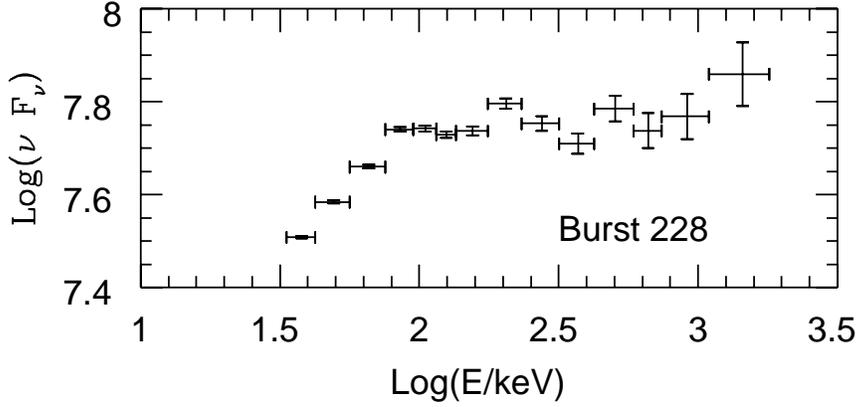

Fig. 2: *The spectrum of burst 228 from the BATSE 2 catalogue.*

**Isotropy:** The observed bursts are distributed isotropically on the sky (see Fig. 3). For 1005 BATSE bursts the observed dipole and quadrupole (corrected to BATSE sky exposure) relative to the galaxy are: $\langle \cos \theta \rangle = 0.017 \pm 0.018$ and $\langle \sin^2 b - 1/3 \rangle = -0.003 \pm 0.009$. This values are, respectively, $0.9\sigma$ and $0.3\sigma$ from complete isotropy [4].

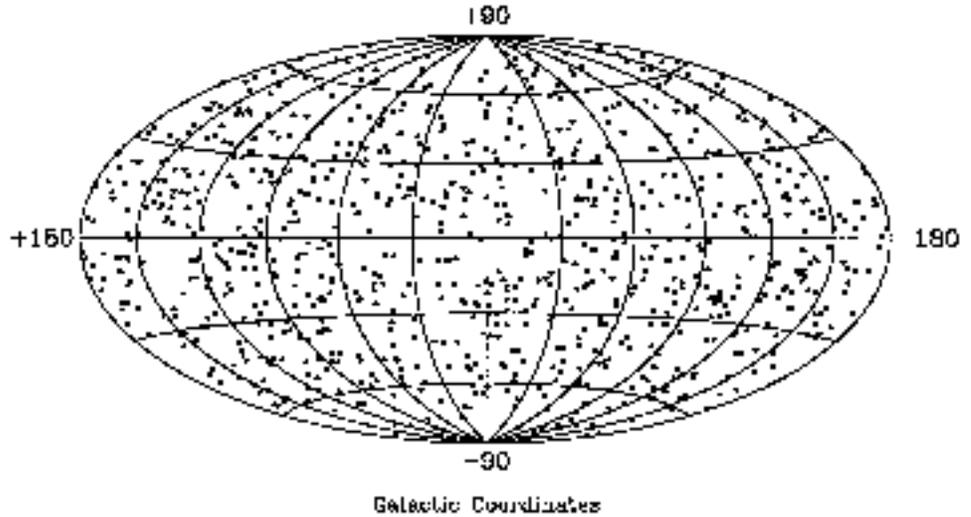

Fig. 3: The Distribution of 1005 bursts on the sky

**Fluence and Flux Distribution:** The limiting fluence observed by BATSE is $\approx 10^{-7}$ergs/cm$^2$. The actual fluence of the strongest bursts is larger by two or three orders of magnitude. A sample of 601 bursts has $\langle V/V_{max}\rangle = .328 \pm 0.012$, which is 14$\sigma$ away from the homogenous flat space value of 0.5 [13]. Correspondingly, the peak count distribution is incompatible with a homogeneous population of sources in Euclidean space. It is compatible, however, with a cosmological distribution (see Fig. 4). Within the cosmological model long bursts are detected by BATSE from $0.2 \approx < z_{min}z < z_{max} \approx 2.1^{+1.}_{-.7}$ (assuming no source evolution). Short bursts are detected from smaller distances [8, 14].

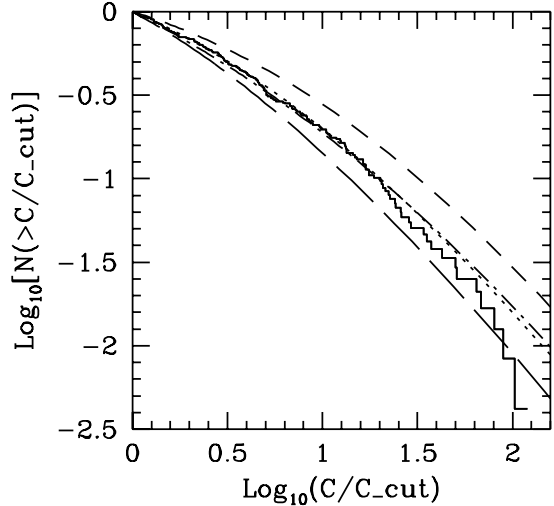

Fig. 4: *The observed long bursts number counts distribution and three theoretical cosmological distributions with $\Omega = 1$, $\Lambda = 0$, $\alpha = -1.5$, standard candles and no source evolution: $z_{max} = 2.1$ (dotted line: best fit), $z_{max} = 1.4$ (long dashed line: lower 1% bound), $z_{max} = 3.1$ (short dashed line: upper 1% bound) and a forth theoretical distribution with $\Omega = 0.1$ and $z_{max} = 2.1$ (dash-dot line).*

**Event Rate:** BATSE observes about one burst per day. With a detection efficiency of $\approx 30\%$ this corresponds to $\approx 1000$ bursts per year. For cosmological sources, with no source evolution, this corresponds to $2.3^{-0.7}_{+1.1} \cdot 10^{-6}$ (long) events per galaxy per year (for $\Omega = 1$ and a galaxy density of $10^{-2}h^3$ Mpc$^{-3}$) [14]. The rate of short bursts is comparable. It goes without saying that if the bursts are beamed with an opening angle $\theta$ than the event rate should increase by a factor of $4\pi/theta^2$ relative to this rate.

**Time Dilation:** Norris *et. al.* [15, 16] found that the dimmest bursts are longer by a factor of $\approx 2.3$ compared to the bright ones. This anti-correlation between the pulse's width and their intensity is compatible with the finding from the count distribution that $z_{max} \approx 2$ and $z_{min} \approx 0.2$ since $(1 + zmax)/(1 + z_{min}) \approx 2.5 \pm .8$. Fenimore & Bloom [17] find, on the other hand, that when the dependence of the duration on the energy band is included, this time dilation corresponds to $z_{max} > 6$ which is incompatible with the count distribution analysis. Clearly, this issue could be resolved only by a combined analysis of the count rate and of the duration using a method that avoids the issue of spectral dependence of the luminosity and the duration [18].

**Soft Gamma-Ray Repeaters (SGRs):** Amongst more than a thousand GRBs there is a unique group of three bursts, including the famous 1989 March 5th event - the strongest GRB ever observed, that are different than all others: (i) Repeated bursts are observed from the same source and (ii) The photon spectrum is distinctly softer. The three SGRs have been identified to coincide with galactic SNRs (the March 5th event coincides with an SNR at the LMC). It is generally accepted that SGRs are different from regular GRBs. However, recently pointed out that the initial part of a SGR spectrum is harder than the rest. This raises the possibility that GRBs and SGRs are more closely related than what was expected before (see for example [19]). I will not explore this possibility in this talk and I will leave it as an observational open question, a very important one, that should be resolved in the future.

**Repetition:** Quashnock and Lamb [20] suggested that there is evidence for repetition of bursts from the same source from the data in the BATSE 1B catalogue. If true this could severely constrain all GRB models. In particular it could rule out the neutron star merger model [21, 22] or any other model based on an 'once in lifetime' catastrophic event. This claim has been refuted by several authors [23, 24] and most notably by the analysis of the 2B data [25]. I have mentioned it here, in spite of that, because of the potential very strong implication of this result, if it is true.

### 0.2.1 Observational Open Questions

There are numerous open observational questions that have not been addressed yet. Most of them deal with finding, yet unknown, correlations between different features of the observed data or classifications of the bursts to sub-classes that show common characteristics. Such relations could help us distinguish between different models. In addition to those, unknown questions, there are several questions concerned with the validity of statements that have been made about the data. The best known among those are:
- *What is the relation, if any, between GRBs and SGRs?*
- *Do GRBs repeat?*
- *Are there absorption lines? or any other spectral features?*
- *Is the time-dilation compatible with the count distribution?*

Some believe that some of these questions have already been answered. The fact that not all agree with that qualifies them as open questions.

## 0.3  A BRIEF SUMMARY

One could say that a fair summary of our present understanding of GRBs can be given in the form of three basic open questions:
- **Where?**
- **What?**
- **How?**

If one is more ambitious one can pose a forth question:
- **Why?**

That this is a reasonable summary is demonstrated by the proliferation of GRB models: a recent review counted more than a hundred. At a time there were more theories than bursts! BATSE has improved this situation enormously: Even the most prolific theoreticians cannot compete with BATSE's rate of one burst per day. Today, in the post-BATSE era, the number of observed bursts exceeds the number of theories by one order of magnitude!

In the rest of the talk I will attempt to show, how does the current data direct us towards some partial answers to those questions and what are the new open questions that have emerged from this understanding. An alternative open question is, of course, to find the flows or the loop-holes in this chain of arguments.

## 0.4  WHERE?

BATSE has revolutionized our ideas about the location of GRBs. Beforehand it was generally believed that GRBs originate in the Galaxy. The isotropy of the sources rules our distant galactic disk population while the incompatibility of the count rate distribution with an Euclidean homogenous distribution rules out local galactic disk sources. The only place left for GRBs at the Galaxy is at the distant parts of an extended galactic halo (with typical distances larger than 100kpc) [26]. On the other hand, the observed distribution is perfectly compatible with a cosmological distribution which is naturally isotropic and homogeneous but with a count distribution that deviates from the $C^{-3/2}$ law due to cosmological effects (see e.g. [27, 28, 29, 14] and other). The cosmological hypothesis is supported by the fact that the predicted [28, 30] anti-correlation between the duration and intensity of the bursts was recently found [15, 16] (see however, [17]). The cosmological interpretation corresponds to an event rate of $2.3^{-0.7}_{+1.1} \cdot 10^{-6}$ events per galaxy per year, We will argue later that at present this is the best (and possibly only) direct clue to the nature of the sources.

It is tempting to enumerate the con and pro arguments for the galactic origin. However, I will not do this for two reasons. First, a *Great Debate* [31, 32] just just took place on this issue and those arguments are discussed extensively there. Second, it is my personal opinion that this is no longer an open question and the present observational data points clearly in the direction of an extra-galactic origin. I will, focus, therefore, on these models, in the rest of my talk. I should stress, however, that current galactic models put the sources so far in the halo that most of the arguments that I present here are valid (with the appropriate numerical scaling) to such sources as well.

In addition to the classical question:
- *Extragalactic or Galactic?*

which both sides believe is not an open question, there are further questions in the context of both models. Some of those are:
- *What are the red-shifts (or distances) from which we observe GRBs?*
- *Can we rule out source evolution in the count distribution analysis?*
- *If the bursts are galactic, then where are the burst from M31?*

## 0.5 HOW?

Before turning to the question *what* can generate GRBs I shall address the question **how** this can be done. Understanding **how** might direct us towards **what**. I shall go backwards from the observations towards the sources and I shall try to keep the discussion as general as possible.

The key to our discussion is the compactness problem: how can a compact source, as inferred from the rapid time variability, emits so much energy in such a short time and remain optically thin, as inferred from the observed non-thermal spectra? The only conventional resolution of this problem known today is extreme-relativistic motion of the source. All other solutions require "new physics". Once we accept the idea that the bursts involve extreme-relativistic motion, it follows that the simplest and energetically most economical way to generate the bursts is via conversion of the kinetic energy of the ultra-relativistic particles to the observed $\gamma$-ray photons. This reduces the question of the origin of GRBs to the questions how to produce large bursts of ultra-relativistic particles and how to convert the kinetic energy of these particles to radiation?

### 0.5.1 The Compactness Problem

The key to understanding GRBs lies, I believe, in understanding how do GRBs bypass the compactness problem. This problem was realized very early on by Schmidt [33]. At that time it was used to show that GRBs can-not originate too far from us. Now, we understand that this interpretation is false and instead we must look for ways to overcome this constraint.

The observed fluence, $F \approx 10^{-7} \mathrm{ergs/cm^2}$, corresponds, for an isotropic source at a distance $D$, to a total energy release of:

$$E = 4\pi D^2 F = 10^{50} \mathrm{ergs} \left(\frac{D}{3000 \mathrm{Mpc}}\right)^2 \left(\frac{F}{10^{-7} \mathrm{ergs/cm^2}}\right), \quad (0.2)$$

Cosmological effects may change this equality by numerical factors of order unity that are not important for our discussion. The rapid temporal variability observed in some bursts (see Fig. 1), $\Delta T \approx 10 \mathrm{msec}$, implies that the sources are compact with a size, $R_i$, smaller than $c\Delta T \approx 3000 \mathrm{km}$. The observed spectrum (see Fig. 2) contains a large fraction (of the order of a few percent) of the $\gamma$-ray photons with energies larger than $2m_e c^2$. I denote by $f_{2m_e c^2}$ this fraction. These photons could interact with lower energy photons and produce electron positron pairs via $\gamma\gamma \to e^+e^-$. The initial optical depth for this process is [34]:

$$\tau_{\gamma\gamma} = \frac{f_{2m_e c^2} \sigma_T F D^2}{R_i^2 m_e c^2}$$

$$10^{13} f_{2m_e c^2} \left(\frac{F}{10^{-7} \mathrm{ergs/cm^2}}\right) \left(\frac{D}{3000 \mathrm{Mpc}}\right)^2 \left(\frac{\Delta T}{10 \mathrm{msec}}\right)^{-2}. \quad (0.3)$$

This optical depth is very large. Even if there are no pairs to begin with they will form rapidly and will Compton scatter lower energy photons. The resulting huge optical depth will prevent us from observing the radiation emitted by the source. Even if the initial spectrum is non-thermal the electron-positron pairs will thermalize it and the resulting spectra will be incompatible with the observations! This is the compactness problem. It is interesting to note that $\tau_{\gamma\gamma} \gg 1$ even if the bursts originate at the extended galactic halo, $D \approx 100 kpc$. Thus, the compactness problem exists and the following analysis is valid even for Galactic halo objects [35]).

It was argued that the only way to avoid the compactness problem is if the sources are nearby ($D < 1 kpc$). At such distances the total energy required is small and equation 0.3 yields $\tau_{\gamma\gamma} \lesssim 1$. Alternatively, it was argued on the basis of this problem that "new physics" is unavoidable if GRBs are at cosmological distances. We will see, however, that it is possible to resolve this paradox within the limits of present day physics.

Compactness would not be a problem if the energy could escapes from the source in some non electromagnetic form which would be converted to electromagnetic radiation at a large distance, $R_X$. This radius will replaces the source's size $R_i < c\Delta T$ in equation 0.3. $R_X$ should be sufficiently large so that $\tau_{\gamma\gamma}(R_X) < 1$. A trivial solution of this kind is a weakly interacting particle, which I will call particle X, which is converted in flight to electromagnetic radiation. The only problem with this solution is that no known particle can play the role of particle X (see however [36]), and this solution requires, indeed, "new physics".

• *Can we rule it out particle X or find a physical candidate?*

### 0.5.2 Relativistic Motion

It is well known that relativistic effects can fool us and, when ignored, lead to wrong conclusions. This has happened here. Consider a source of radiation that is moving towards an observer at rest with a relativistic velocity characterized by a Lorentz factor, $\Gamma = 1/\sqrt{1 - v^2/c^2} \gg 1$. The observer detects photons with energy $h\nu_{obs}$. These photons have been blue shifted and their energy at the source was $\approx h\nu_{obs}/\Gamma$. Fewer electron will have energies larger than $2m_e c^2$ and the fraction $f_{2m_e c^2}$ at the source is smaller by a factor $\Gamma^{-\alpha}$ than the observed fraction. At the same time relativistic effects require now: $R_i < \Gamma^2 c \Delta T$. The radius from which the radiation is emitted could be larger than the original estimate by a factor of $\Gamma^2$.

$$\tau_{\gamma\gamma} = \frac{f_{2m_e c^2}}{\Gamma^\alpha} \frac{\sigma_T F D^2}{R_i^2 m_e c^2}$$

$$\approx \frac{10^{13}}{\Gamma^{(4+\alpha)}} f_{2m_e c^2} \left(\frac{F}{10^{-7} \text{ergs/cm}^2}\right) \left(\frac{D}{3000 \text{Mpc}}\right)^2 \left(\frac{\Delta T}{10 \text{msec}}\right)^{-2}, \quad (0.4)$$

where the relativistic limit on $R_i$ was included in the second line. The compactness problem can be resolved if the sources are moving relativistically towards us with Lorentz factors $\Gamma > 10^{13/(4+\alpha)} \approx 10^2$. A more detailed discussion [?] gives practically the same result. Such extreme-relativistic motion ($v \approx 0.9995c$) was never detected (or even suspected to exist) for a macroscopic object in the Universe! This resolution of the paradox is clearly within conventional physics, as all that it requires is special relativistic effects. But it involves extremely relativistic motion that were never met before.

The potential of relativistic motion to resolve the compactness problem was realized in the eighties by Goodman [37], Paczyński [38] and Krolik and Pier [39]. There was however a drastic difference between the first two approaches and the last one. Goodman [37] and Paczyński [38] considered relativistic motion in the dynamical context of fireballs. In this case the relativistic motion is an integral part of the burst mechanism. Krolik and Pier [39] considered, on the other hand, a kinematical solution, in which the sources move relativistically and this motion is not necessarily related to the mechanism that produces the bursts.

Is the kinematic scenario feasible? In this scenario the source moves relativistically as a whole. The radiation is beamed with an opening angle of $\Gamma^{-2}$. The total emitted energy is smaller by a factor $\Gamma^{-3}$ than the isotropic estimate given in equation 0.2. The total energy required, however, is at least $(Mc^2 + 4\pi FD^2/\Gamma^3)\Gamma$, where $M$ is the rest mass of the source (the energy would be larger by an additional amount $E_{th}\Gamma$ if an internal energy, $E_{th}$, remains in the source after the burst has been emitted). The whole process becomes very wasteful if the kinetic energy, $Mc^2\Gamma$ is much larger than the observed energy of the burst, $(4\pi/\Gamma^2)FD^2$.

One can find several arguments that show that in most reasonable cases this is exactly what happens and the total required energy is so large that the model becomes unfeasible. The only exception is the most energetically) economical situation when the kinetic energy itself is the source of the observed radiation. This is also the most conceptually economical situation, since in this case the $\gamma$-ray emission and the relativistic motion of the source are related and are not two independent phenomena. This will be the case if GRBs result from the slowing down of ultra relativistic matter. This idea was suggested by Mészaros, & Rees [40, 41] in the context of slowing down of fireball accelerated material [43] by the ISM and by Narayan *et. al.* [44] and independently by Mészaros, & Rees [45] in the context of self interaction and internal shocks within the fireball. However, it seems to be much more general and in my mind it is an essential part of any GRB model regardless of the acceleration mechanism of the relativistic particles!

Assuming that GRBs result from slowing down of relativistic bulk motion of massive particle we find that the required mass of the ultra-relativistic source is:
$$M = \frac{\theta^2 FD^2}{\Gamma \epsilon_c}, \qquad (0.5)$$
where $\epsilon_c$ is the conversion efficiency and $\theta$ is the opening angle of the emitted radiation. The relativistic motion does not imply relativistic beaming as is sometimes mistakenly believed. $\theta$ can be as small as $\Gamma^{-1}$, the limiting relativistic beaming factor, if the matter has been accelerated along a very narrow beam. Notice that relativistic beaming requires an event rate larger by a ratio $4\pi\Gamma^2$ compared to the observed rate. With observation of about one burst per $10^{-6}$ year per galaxy this imply one event per year per galaxy! $\theta^2$ can be as large as $4\pi$ as would be the case if the motion results from the relativistic expansion of a fireball [37, 38]. The opening angle can also have any intermediate value if it emerges from a beam with an opening angle $\theta > \Gamma^{-1}$, as will be the case if the source is an anisotropic fireball [46, 47] (see Fig. 5) or an electromagnetic accelerator with a modest beam width. In the last two cases each observer will see, indeed, radiation beamed towards him or her from a region whose width is $\Gamma^{-1}$. However, observes that are more than $\Gamma^{-1}$ apart will still see the a burst from the same source.

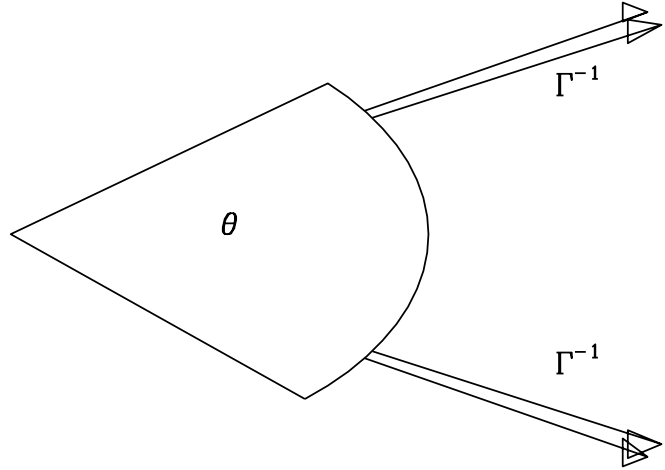

*Fig. 5: Radiation from a relativistic bean with a width $\theta$. Each observer will detect radiation only from a very narrow beam with a width $\Gamma^{-1}$. The overall angular size of the observed phenomenon can vary, however, with*
$$\Gamma^{-2} < \theta^2 < 4\pi.$$

It is somewhat amusing that we have found particle X in the simple form of a proton. These protons escape from the source carrying the energy as kinetic energy. To produce a GRB they should convert their kinetic energy to radiation, otherwise they are useless. The next question is, therefore, how is the energy converted?

### 0.5.3 Slowing Down of Relativistic Particles

The cross section for a direct nuclear or electromagnetic interaction between the relativistic protons and the ISM protons is far too small to convert efficiently the kinetic energy to radiation. The only way that the protons can be slowed down in via some collective interaction such as a collisionless shock. The existence of supernova remnants in which the supernova ejects is slowed down by the ISM indicates that collisionless shocks do form in similar circumstances [40].

GRBs are the relativistic analogues of SNRs. In both cases the phenomenon results from the conversion of the kinetic energy of the ejected matter to radiation. Even the total energy involved is comparable. The crucial difference is the amount of ejected mass. SNRs involve several solar masses. The corresponding velocity is several tens of thousands of second, much less than the speed of light. The interaction that takes place on scales of several pc is observed for thousands of years. In GRBs the masses are smaller by several orders of magnitude and with the same energy the matter attains ultra-relativistic velocities. The interaction with the ISM takes place on a comparable but slightly smaller distance scale. Special relativistic effects reduce, however, the observed duration of the bursts to a few seconds!

The exact details of the microscopic processes that take place in the shocks are still an open question. However, shocks are independent of microscopic physics and we can safely examine the global conditions that arise in the shocks without this information. I will present here, first a simple calculation of relativistic plastic collision which provides a guide line, for what happens in the shocks. I discuss next internal shocks and finally external shocks. I show that in all cases the simple analysis fails and there are interesting open questions in all scenarios[1].

**Relativistic Plastic Collisions**

Consider a pulse of ultra-relativistic particles with a total mass, $M$, and a Lorentz factor $\Gamma$ that collides with an external mass $m$ that moves with a Lorentz factor $\gamma$ in the same direction. After the collision both masses move at the same velocity, with a Lorentz factor $\Gamma_f$. Energy and momentum conservation yield:

$$\begin{aligned} M\Gamma + m\gamma &= (M + m + \mathcal{E}/c^2)\Gamma_f \\ M\sqrt{\Gamma^2 - 1} + m\sqrt{\gamma^2 - 1} &= (M + m + \mathcal{E}/c^2)\sqrt{\Gamma_f^2 - 1}, \end{aligned} \quad (0.6)$$

where $\mathcal{E}$ is the internal energy generated in the collision. To reach effective conversion of kinetic to thermal energy we require $\mathcal{E} \approx Mc^2/2$.

There are two interesting limits to this set of equations. The first, corresponds to an internal collision between shells that are moving at a comparable but different velocities. In this case $\gamma \lesssim \Gamma$. The motion of one shell with respect to the other is only mildly relativistic: with a Lorentz factor of $\sqrt{\Gamma/\gamma} \approx \sqrt{2}$. For efficient energy conversion we need $m \approx M$ as intuitively expected.

---

[1] The following discussion is somewhat more technical than the rest of this lecture. The uninterested reader can skip to the open questions subsection at the end of this section.

The second limit is when a shell collides with an external matter, such as the ISM, which is at rest $\gamma \approx 1$. In this case:

$$m \approx M/\Gamma \ll M \qquad (0.7)$$

is needed to yield $\mathcal{E} \approx M/c^2 2$. The surprising and non intuitive result is that the external mass needed to slow down the matter and to extract half of the kinetic energy is smaller than the original mass by a factor of $\Gamma$ [40].

**Compactness Revisited**

The kinetic energy is converted to radiation in large enough radii in which the system is optically thin. Additionally the energy of the photons is lower than the observed energy by the relevant Lorentz factor. Both effects lead to a nice resolution of the compactness problem.

An inner shell moving at $\Gamma$ overtakes an outer shell moving at $\Gamma/2$ at:

$$R_c \approx \Gamma^2 \delta R \approx 10^{12} \text{cm} \left(\frac{\delta R}{1000 \text{km}}\right) \left(\frac{\Gamma}{100}\right)^{-2} \qquad (0.8)$$

where $\delta R$ is the initial separation between the shells in the observer's rest frame. Substituting $R_c$ in equation 0.4 we find that $R_c$ is large enough so that $\tau \approx 1$, the region is optically thin and photons escape easily to infinity. The observed time scale for the collision is:

$$\Delta T_{obs} \approx R_c/(\Gamma^2 c) \approx \frac{\delta R}{c}. \qquad (0.9)$$

$\Delta T_{obs}$ is small enough to produce even the fastest observed variation if $\delta R$ is smaller than $10^8$cm. The total duration of the bursts in this scenario is simply the duration of the emitted pulse of relativistic particles.

In our second scenario the ejected matter is slowed down by the interaction with the ISM. Our simple example shows that this happens at $R_\Gamma$, where the shell collects a mass $m \approx M/\Gamma$:

$$R_\Gamma = \left(\frac{M}{(4\pi/3)n_{ism}}\right)^{1/3} = \left(\frac{3FD^2}{\epsilon_c m_p c^2 n_{ism} \Gamma}\right)^{1/3} =$$

$$6 \cdot 10^{16} \text{ cm} \epsilon_c^{-1/3} \left(\frac{F}{10^{-7} \text{ergs/cm}^2}\right)^{1/3} \left(\frac{D}{3000 \text{Mpc}}\right)^{2/3} \left(\frac{\Gamma}{100}\right)^{-1/3}, \qquad (0.10)$$

where $n_{ism}$ is the ISM number density ($n_{ism} \approx 1$ particle/cm$^3$) and we have used the second equality equation 0.5 that relates the ejected mass to the observed fluence and the distance to the source. The distance, $R_\Gamma$, is, incidentally, independent of the opening angel $\theta$ of the beam. If $\theta$ is smaller less mass is needed, but correspondingly less mass is dragged from the interstellar medium. Substitution of $R_\Gamma$ in equation 0.4 reveals that the optical depth is much smaller than one.

The observed time scale of the burst is now:

$$\Delta T_{obs} \approx R_\Gamma/(\Gamma^2 c) \approx 200 \text{ sec} \epsilon_c^{-1/3} \left(\frac{F}{10^{-7}\text{ergs/cm}^2}\right)^{1/3} \times$$

$$\left(\frac{D}{3000\text{Mpc}}\right)^{2/3} \left(\frac{n_{ism}}{1 \text{ cm}^{-3}}\right)^{-1/3} \left(\frac{\Gamma}{100}\right)^{-7/3}. \qquad (0.11)$$

This value is comparable to the duration of the long bursts. It is very sensitive to $\Gamma$. An increase of $\Gamma$ by a factor of 10 will reduce the time scale by two orders of magnitudes to the transition regime between long and short bursts. Another increase by a factor of 10 in $\Gamma$ is required to reach the rapid variability observed in some of the short bursts. However, as we will see shortly, there is another time scale in this scenario which is generally shorter and which could determine this variability.

**Shock Conditions**

We consider a cold shell (whose internal energy is negligible compared to the rest mass energy) that overtakes another cold shell or moves into the cold ISM. Two shocks form: an outgoing shock that propagates into the ISM or into the external shell and a reverse shock that propagates into the inner shell, with a contact discontinuity between the shocked material (see Fig. 6). Two quantities determine the shocks' structure: $\Gamma$, the Lorentz factor of the motion of the inner shell (denoted 4) relative to the external matter (denoted 1) and the ratio between the particle number densities in these regions, $f \equiv n_4/n_1$. I ignore here a third quantity, the adiabatic index of the matter, which gives rise only to factors of order unity.

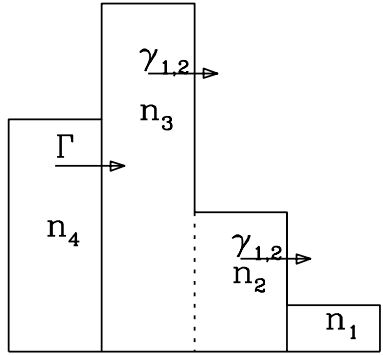

*Fig. 6: Schematic density profile in the interaction between a relativistic shell of matter (region 4) and the ISM (region 1). The shocked regions are 2 - shocked ISM material)and 3 - shocked shell material. The forward shock is marked by a solid line. The reverse shock by a dotted line and the contact discontinuity between regions 2 and 3 is marked by a dashed line.*

In the original analysis of [40, 41, 48] it was assumed that both shocks are relativistic. In fact this takes place only if $f < \Gamma^2$. If this condition holds and if $\Gamma \gg 1$ then the shock equations between regions 1 and 2 yield: [49, 50, 46]:

$$\gamma_{1,2} = f^{1/4}\Gamma^{1/2}/\sqrt{2} \ ; \ n_2 = 4\gamma_{1,2}n_1 \ ; \ e \equiv e_2 = \gamma_{1,2}^2 n_1 m_p c^2, \quad (0.12)$$

where $\gamma_{1,2}$ is the Lorentz factor of the motion of the shocked fluid relative to the rest frame of the fluid at 1 (an external observer for interaction with the ISM and the outer shell in case of internal collision). The Lorentz factor of the shock front itself is $\sqrt{2}\gamma_{1,2}$. Similar relations hold for the reverse shock:

$$\gamma_{3,4} = f^{-1/4}\Gamma^{1/2}/\sqrt{2} \ ; \ n_3 = 4\gamma_{3,4}n_4, \ ; \ e_3 = e. \quad (0.13)$$

The last condition follows from the equality of pressures on the contact discontinuity.

If $f < \Gamma^2$ the reverse shock is non-relativistic and:

$$\gamma_{1,2} \approx \Gamma \ ; \quad \gamma_{3,4} \approx 1. \quad (0.14)$$

$$n_2 \approx 4\Gamma n_1, \quad ; \quad e \equiv e_2 = 4\Gamma^2 n_1 m_p c^2 \quad ; \quad n_3 = 7n_4, \quad ; \quad e_3 = e. \qquad (0.15)$$

Comparable amounts of energy are converted to thermal energy in both shocks when both shocks are relativistic. But only a negligible amount of energy is converted to thermal energy in the reverse shock if it is Newtonian [51]. The above shock conditions follow from a planar analysis. However, numerical simulations of spherical ultra-relativistic shocks [51] show that these conditions are valid at each moment of time even for spherical systems.

### A Model for The Observed Spectra: Synchrotron Emission from the Shocked Regions

We turn now to a toy model calculations of the observed spectra of the photons emitted form the shocked regions. The shock conditions depend on energy and momentum conservations which are robust and independent of the details of the viscosity and other microscopic mechanisms. The radiation mechanism depends, on the other hand, heavily on those details. Hence the following discussion should be considered only as an example. Following [41, 42] we have chosen the synchrotron mechanism, which is a classical source of non-thermal radiation.

I assume equipartition between different energy densities. Thus, all energy densities can be expressed in terms of the thermal energy density of the protons, $e$. Equipartition between magnetic and thermal energies yields:

$$B^2/4\pi = \epsilon_B e = \epsilon_B \gamma_{1,2}^2 n_1 m_p c^2, \qquad (0.16)$$

where $\epsilon_B$ measures the deviation from equipartition. There is no index to $B$ since the energy densities in 2 and 3 are the same, and from this follows the equality of the magnetic fields.

Equipartition between the energy density of the electrons and the protons: $e_{eln} = \epsilon_{ep} e$ yields a typical Lorentz factor, $\gamma_{2e}$, of the thermal motion of the electrons that is larger by $(m_p/m_e)\epsilon_{ep}$ than the Lorentz factor of the thermal motion of the protons, $\approx \sqrt{2}\gamma_{1,2}$. The typical energy of a photon emitted by the synchrotron process is:

$$h\nu_{synch} = h\gamma_{2e}^2 \frac{eB}{m_e c} \approx \left(\frac{m_p}{m_e}\right)^2 \frac{\sqrt{4\pi}\sqrt{m_p} he}{m_e} \epsilon_{ep}^2 \epsilon_B^{1/2} \gamma_{1,2}^3 n_1^{1/2}. \qquad (0.17)$$

The emitted energy is blue shifted by a factor of $\gamma_{1,2}$ relative to an observer at the frame 1 and by another factor of $\gamma_1$ if this frame is moving relative to an observer at rest at infinity. Thus, the observed energy is:

$$h\nu_{syn} \approx \left(\frac{m_p}{m_e}\right)^2 \frac{\sqrt{4\pi}\sqrt{m_p} he}{m_e} \epsilon_{ep}^2 \epsilon_B^{1/2} \gamma_{1,2}^4 \gamma_1 n_1^{1/2} = 0.01 \text{ eV } \epsilon_{ep}^2 \epsilon_B^{1/2} \gamma_{1,2}^4 \gamma_1 n_1^{1/2}. \qquad (0.18)$$

A similar estimate yields the same expression divided by f for the typical energy of a synchrotron photon emitted by a relativistic reverse shock.

**Internal Shocks**

Internal shock take place when an inner shell overtakes an outer shell. There are several complications in this scenario. First, the relative Lorentz factor, $\sqrt{\Gamma/\gamma}$ is not significantly larger than unity. The shocks are at best mildly relativistic and equations 0.12 and 0.13 are not valid. The density ratio, $f$ is also of order unity and since both densities decrease like $R^{-2}$ it remains constant in time. However, since both $f$ and $\Gamma$ are of order unity a relatively small variation of $f$, can cause one of the shocks to be non relativistic. Finally, Waxman and Piran [52] have shown that shell crossing is, quite generally, unstable. It is not known yet what are the full implication of this instability.

In spite of all those limitations consider as an example, the conditions in an internal shock for a specific case of a large relative Lorentz factor, $sqrt\Gamma/\gamma = 4$, and equal densities $f = 1$. Using equations 0.12 and 0.13 we find:

$$\gamma_{1,2} = \gamma_{3,4} \approx \sqrt{2} \ ; \ n_2 = n_3 \approx 4\sqrt{2}n_1, \ ; \ e \approx 8n_1 m_p c^2. \quad (0.19)$$

The observed energy of emitted photons is:

$$h\nu \approx 10 \text{ GeV } \epsilon_{ep}^2 \epsilon_B^{1/2} \epsilon_c^{-1/2} \left(\frac{E}{10^{50}\text{ergs}}\right)^{1/2} \left(\frac{\Gamma}{100}\right)^{-3/2} \left(\frac{\Delta R}{10^3 \text{km}}\right)^{-3/2}, \quad (0.20)$$

where we have used equations 0.5 and 0.8 to estimate the density, $n_1$, at the time of shell crossing. The conditions at region 3 are similar and the emitted photons have the same energy. The resulting energy is high, which is a good sign. But it is too high and it is not in the right energy range. This might be resolved if $\Gamma$ or $\Delta R$ are larger or if the various equipartition factors are smaller. Alternatively, these shocks might provide the observed GeV photons, while the interaction with the ISM might provide the rest of the burst.

**Shocks with the ISM**

Two new phenomena appear when we consider the interaction of a relativistic shell with the ISM. First, the density ratio between the relativistic shell and the ISM, $f$, is initially so large that the reverse shock is Newtonian. The factor, $f$, decreases with time, as the shell's density is proportional to $R^{-2}$ while the ISM density remains constant. However, in most cases the reverse shock remains Newtonian until the kinetic energy is converted into thermal energy. Secondly, the reverse shock may reach the inner edge of the shell before $R_\Gamma$. At this stage a reflected rarefraction wave begins to move forwards. This wave is, in turn, reflected from the contact discontinuity, between the shell's material and the ISM material and another reverse shock begins. A series of weak reverse shocks and rarefraction waves create quickly a self similar profile that describes well the shell's material while most of the action takes place in the forward shock [51] There are two critical radii: $R_\Delta = R_\Gamma/\sqrt{\xi}$, the radius where the reverse shock reaches the inner boundary of the shell, and $R_N = \xi R_\Gamma$, the radius where the reverse shock becomes relativistic. There are two possible scenarios depending on which radius is larger. The dimensionless ratio:

$$\xi \equiv \frac{E^{1/6}}{\epsilon_c^{1/6}(m_p c^2)^{1/6}\Delta R^{1/2} n_{ism}^{1/6} \Gamma^{4/3}} =$$

$$150 \epsilon_c^{-1.6} \left(\frac{E}{10^{50}\text{ergs}}\right)^{1/6} \left(\frac{n_{ism}}{1\text{ cm}^{-3}}\right)^{-1/6} \left(\frac{\Delta R}{10^7\text{cm}}\right)^{-1/2} \left(\frac{\Gamma}{100}\right)^{-4/3}, \quad (0.21)$$

determines which one prevails [51].

For our canonical parameters $\xi > 1$. In this case

$$R_\Delta < R_\Gamma < R_N, \quad (0.22)$$

and reverse shock is Newtonian (see Fig. 7a). The time for crossing the shell is relatively short. Therefore the whole mass of the shell participates in the energy conversion and $R_\Gamma$ is a good estimate for the radius where the kinetic energy is converted to thermal energy. Most of the energy is emitted from the forward shock region (region 2) which is extremely hot since $\gamma_{1,2} \approx \Gamma$. Using equation 0.18 we find that the typical energy of synchrotron photons emitted from this region is:

$$h\nu \approx 1 \text{ MeV} \epsilon_{ep}^2 \epsilon_B^{1/2} \left(\frac{\Gamma}{100}\right)^4 \left(\frac{n_{ism}}{1\text{ cm}^{-3}}\right)^{1/2}. \quad (0.23)$$

This energy seems right where it should be. However, the dependence on the forth power of $\Gamma$ suggests that this might be nothing more than a nice coincidence. Since the energy is emitted at $R_\Gamma$ equation 0.11 provides a good estimate for the duration of the bursts. In addition we find that the observed time for crossing the shell, $R_\Delta/(\Gamma^2 c) \approx R_\Gamma/(\Gamma^2 c) = \Delta T_{obs}/\sqrt{\xi}$ gives a reasonable scale for the variability of the bursts [51].

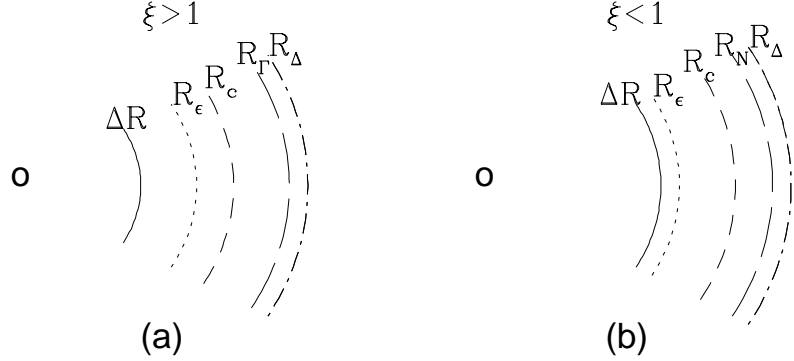

*Fig. 7: (a) Schematic description of the different radii for the case $\xi > 1$. The different distances are marked on a logarithmic scale. Beginning from the inside we have $\Delta R$ the initial size of the shell, $R_\eta$ the radius in which a fireball becomes matter dominates (see the following discussion), $R_c$, the radius where inner shells overtake each other and collide, $R_\Delta$ where the reverse shock reaches the inner boundary of the shell and $R_\Gamma$ where the kinetic energy of the shell is converted into thermal energy. (b) Same as (a) for $\xi < 1$. $R_\Gamma$ does not appear here since it is not relevant. $R_N$ marks the place where the reverse shock becomes relativistic.*

The situation is drastically different if both $\Gamma$ and $\Delta R$ are large enough so that $\xi < 1$ (see Fig. 7b). Now

$$R_N < R_\Gamma < R_\Delta. \tag{0.24}$$

The forward shock become relativistic early on. From this moment onwards the conversion of kinetic energy to thermal energy is very efficient. However, now only a small fraction of the shell is shocked by the time that $R_\Gamma$ is reached. A significant fraction of the kinetic energy is converted to thermal energy only at $R \approx R_\Delta$ i.e. when a significant fraction of the shell's material is shocked and heated. Since $R_\Delta > R_\Gamma$ the region is optically thin. The time scale, which is now $R_\Delta/(\Gamma^2 c) \approx 1.7 \text{sec } (\Delta R/10^{10}\text{cm})^{1/4}(\Gamma/1000)^{-5/3}$. Note that we have used different "canonical" numbers in this last equation. The resulting synchrotron energy from the forwards shock is:

$$h\nu \approx 100\text{MeV } \epsilon_{ep}^2 \epsilon_B^{1/2} \epsilon_c^{-1/2} \left(\frac{E}{10^{50}\text{ergs}}\right)^{1/2} \left(\frac{\Delta R}{10^{10}\text{cm}}\right)^{-3/2}. \tag{0.25}$$

Comparable amount of energy is emitted now from the reverse shock with a typical energy of:

$$h\nu \approx 10\text{keV} \ \epsilon_{ep}^2 \epsilon_B^{1/2} \left(\frac{\Gamma}{1000}\right)^2 \left(\frac{n_{ism}}{1 \text{ cm}^{-3}}\right)^{1/2}. \tag{0.26}$$

The typical energy from the forward shock is not independant of $\Gamma$ but it is slightly too high. The typical energy from the reverse shock is almost in the right range but again there is rather strong dependence on $\Gamma$.

**Open Questions**

This long section contained some detailed analysis of the conditions at the place where the kinetic energy is converted to thermal energy and where the radiation is emitted. We have seen that internal shocks or shocks with the ISM can convert the kinetic energy of the relativistic baryons to radiation and that the process can take place with the right time scale. While this analysis shows that we possibly understand the main framework it is clear that many details are missing. The basic flow of this analysis is that there is no mechanism that anchors the resulting photons to the observed range of low-energy $\gamma$-rays. We find that the observed energy is quite strongly dependent on $\Gamma$ and unless some robust process keeps all quantities that appear in equations 0.20 0.23 0.25 and 0.26 roughly constant we should observe similar events with uv/x-ray photons, on one hand and much harder $\gamma$-rays on the other hand. Such events are not observed! The fact that there is no clear explanation for the most basic feature of GRBs: their unique spectral range, may cause the reader to question the whole analysis. He or she might be right, but in fairness one should recall that almost none of the GRB models that have been suggested so far provides a satisfactory answer to this question. The most puzzling question is, therefore:
• *What nails the observed spectrum to the soft $\gamma$-ray band? - or why don't we see comparable events in other parts of the spectrum?*
A related question is:
• *Why there are no counterparts to GRBs events at other parts of the spectrum?*
Two other questions that are emerge from this analysis are:
•*What are the relative roles of internal vs. ISM shocks?*
•*Can we explain the bimodal distribution of durations in terms of internal vs. ISM shocks or in terms of $\xi > 1$ vs. $\xi < 1$ shocks?* When we recall that both internal shocks and an interaction with the ISM can take place in the same burst we realize that the radiation from the internal shocks will arrive at a time interval $\approx R_\Gamma/\Gamma^2 c$ or $R_\Delta/Gamma^2 c$ before the radiation from the interaction with the ISM. With reasonable parameters this interval would be several tens of seconds which leads us to another question:

• *Can internal shocks produce the precursors observed in some bursts?*

I should remark now on the numerical values of the parameters used in this analysis. We have seen early on that equation 0.4 gave a lower limit on $\Gamma$ which was of order 100. Being conservative I have used this lower limit as the canonical value in this analysis. Historically, one would have used $\delta R \approx \Delta R \approx 1000$ km as indicated by the shortest time scale variability observed in some of the bursts. However, we have seen now that in fact this value is not necessarily relevant any more. Now, different scenarios put different constraints on the width of the shell - or alternatively on the size of the internal engine. Internal shocks seem to require long pulses (with a duration comparable to the observed duration) and variability on scale of $\approx 10^8$cm. ISM shocks seem to require narrow bursts, with a total width of less than $10^7$cm. Lorentz factors, $\Gamma \approx 10^4$ can increase, however, the allowed $\Delta R$ up to $10^{13}$cm. While both models indicate the need for a compact source the situation is not clear and the immediate question that follows is:

•*Can we determine $\Gamma$ and $\Delta R$ from the current observations?*

### 0.5.4 The Acceleration Mechanism?

We have seen that GRBs require a short burst of ultra-relativistic particles, with a total energy of $\approx 10^{50}/4\pi$ ergs per steradian and a Lorentz factor of a hundred or larger. According to the analysis presented so far, the observed $\gamma$-rays are produces when the ultra-relativistic particles are slowed down. However, there are no direct observations that can tell us about the acceleration phase. This brings us directly to the next open question:

•*What is the Acceleration Mechanism?*

There are two clear alternatives: A non-thermal, most likely electromagnetic, mechanism or a thermal mechanism, in which the particles are accelerated by thermal pressure. This second mechanism falls under the general category of fireballs.

There is little that I can say about the non-thermal acceleration mechanism. The analogy to pulsars and other steady state sources that produce high energy radiation and that accelerate particles to relativistic velocities is appealing. One has to remember, however, that the energies reached here are significantly larger than those observed in other astrophysical jets. The observed asymmetry in the temporal pulse profiles (rapid rise vs. slower fall [56]) practically rules out a "light house" (i.e. a rotating beam) model, which will be symmetric in the initial rise and the final decay, as an alter-native to pulsed beams. Even with relativistic beaming the energy requirements from this accelerator are quite severe: $\approx 10^{46}/(\Gamma/100)^2$ ergs within a few second! This brings us immediately to a mysterious open question:
•*Is there a suitable non-thermal (electromagnetic) acceleration mechanism that satisfies these constraints?*

I will focus in the rest of this section on thermal acceleration that is on the fireball process.

**Fireballs**

GRBs involve the release of $\approx 10^{50}$ ergs of radiation into a small volume with a radius of $\approx 10^3$ km. Equation 0.3 shows that such a system will be optically thick to $\gamma\gamma \rightarrow e^+e^-$. The radiation will not be able to escape and the large optical depth will cause it to reach thermal equilibrium rapidly, with a temperature: $T_i \approx 1\text{MeV}(E/10^{50}\text{ergs})^{1/4}(R_i/10^3\text{km})^{-3/4}$. At this temperature there is a copious number of $e^+ - e^-$ pairs that contribute to the opacity via Compton scattering. The system turns into a fireball: a dense radiation and electron-positron pairs fluid. The fluid expands under its own pressure and it cools adiabatically due to this expansion. The phase ends when the fireball becomes optically thin, and stops behaving like a fluid. The question where does this happens and what follows depend critically on the fireball's constitution.

Consider, first, a pure radiation fireball. This fireball expands and cools with $T \propto R^{-1}$. The electron-positrons gradually annihilate and disappear. The phase ends when the local temperature is $\approx 20$ KeV, and sufficiently few high energy photons are left so that $\tau \approx 1$ [37, 43]. The photons escape freely as the fireball becomes transparent. In the meantime the fireball has been accelerated by its own pressure and the radiation-electron-positron fluid has reached extreme-relativistic velocities [37, 38] with $\Gamma \approx R/R_i \approx T_i/T$. The observed photon energy, as seen by an observer at infinity, is now blue shifted practically back to the original temperature: $T_{obs} \approx \Gamma T \approx T_i$. The resulting photon spectrum is, however, thermal, and very different from those observed in GRBs.

Astrophysical fireballs may include baryonic matter in addition to radiation and $e^+e^-$ pairs. The baryons affect the fireball in two ways: The electrons associated with the baryons increase the opacity and delay the escape of radiation. The baryons are also dragged by the accelerated leptons and this requires a conversion of the radiation energy into a kinetic energy. The second effect is more important and we will focus on it here (see [43, 54, 46, 55] for a more detailed discussion of fireball evolution). The acceleration of the baryons leads to a transition from the initial radiation dominated phase (in which most of the energy is in the form of radiation) to a matter dominated phase (in which the energy is mostly the kinetic energy of the baryons). The factor $\eta \equiv E_i/Mc^2$, the ratio of the initial radiation energy $E_i$ to the rest energy $Mc^2$ determines the location of the transitions and the fate of the fireball. The transition takes place at:

$$R_\eta = 2R_i\eta = 2 \times 10^{10} \text{cm} \left(\frac{R_i}{1000\text{km}}\right)\left(\frac{\eta}{100}\right). \qquad (0.27)$$

The overall outcome of the fireball is the same as the outcome of a pure radiation fireball if $R_\eta$ is larger than the radius in which the fireball becomes optically thin. In this case all the initial energy is still carried away by photons, with a thermal spectrum. For most reasonable baryonic loads $R_\eta$ is, however, smaller than the radius in which the fireball becomes optically thin. Such a fireball results in a relativistic expanding shell of baryons, whose kinetic energy equals the total initial energy [43, 53]. Energy conservation dictates that

$$M = \frac{E_i}{c^2\Gamma} = 5 \cdot 10^{-7} m_\odot \left(\frac{E}{10^{50}\text{ergs}}\right)\left(\frac{\Gamma}{100}\right)^{-1}, \qquad (0.28)$$

where $E_i$ is the total *initial* energy (and not the observed energy). Comparison with the definition of $\eta$ reveals that $\Gamma \approx \eta$. The width of the shell is comparable to the original size of the fireball: $\Delta R \approx R_i$ [54] . Surprisingly, we discover that the most likely outcome of a fireball is just what we need: an narrow relativistic shell of baryons with a very large $\Gamma$.

We have estimated in equation 0.28 the mass load for a spherical fireball. It goes without saying that $E_i$ and $M$ are smaller by a factor $\theta^2/4\pi$ if the fireball is not spherical and has an angular opening $\theta$. A quick glance at this mass limits reveals that the baryonic load of the fireball must be extremely small, otherwise the motion will not be relativistic. This leads us immediately to the another open question:
• *How can one produces "clean" enough fireballs with sufficiently small baryonic loads?*
Some have argued that this is impossible and used this condition as an argument against the thermal acceleration mechanism. Others have argued that one can use this constraint to rule out specific models for the "engine" that generates the fireball as some engines cannot produce "clean" enough fireballs. My personal view is that this is still an open question, a very puzzling one.

Before leaving this topic, it is worth mentioning that fireballs are not necessarily spherical as their name imply. The equations governing a spherical fireball are simplest to derive. However, it has been shown that even a small fraction of a spherical shell, that is any beam whose width $\theta > \Gamma^{-1}$, behaves locally as if it is a part of a spherical shell [47]. Thus, fireball, could in fact be firebeams, if sufficient collimation takes place at the initial stages (see Fig. 5).

## 0.6 WHAT?

We now turn to our third question, **what?** I address this question after discussing **where?** and **how?** because I hope that the previous discussion have constrained the sources that could produce GRBs. I will summarize first what have we learnt so far on the nature of the "engines" and then will turn to astrophysical models.

### 0.6.1 What do We Need from the Internal Engine?

GRBs are produces by some internal "engine" that supplies the energy for the process. This "engine" is well hidden from direct observations and it will be difficult to determine what it is from the available data. We have concluded that this "engine" should supply us with a short pulse of extreme relativistic particle. The engine should accelerate $4 \cdot 10^{-8}/(\Gamma/100) m_\odot$ per steradian to $\Gamma > 100$. The minimal total energy required (assuming full relativistic beaming) is $10^{46}/(\Gamma/100)^2$ ergs. The maximal mass allowed is $5 \cdot 10^{-7} m_\odot/(\Gamma/100)$ (assuming a spherical burst). The total duration of the pulse varies from several msecs to several hundred seconds. The size of the source is, most likely, less than 1000km. The acceleration can be direct, via an (unknown yet) non-thermal (most likely electromagnetic) process or indirectly via a fireball phase. The source of the fireball should produce high energy photons with a total energy of $8 \cdot 10^{48}$ ergs per steradian, with no more than $4 \cdot 10^{-8} m_\odot$ per steradian within this radiation.

These are the only constraints on the sources of GRBs. These constrains are indirect and follow from our analysis. The compactness problem tells us that it is impossible to observe the sources directly, at least with electromagnetic radiation, and hence there are no direct observational constraints. The only direct observational constraint is the rate of the bursts:$\approx$ 1 per $10^6$ years per galaxy for isotropic bursts. However, even this limit is not strict as an uncertainty in the beaming angle, $\theta$, of the bursts leads to an uncertainty of order $4\pi/\theta^2$ in the rate. Any process that satisfies these constraints, and whose event rate is compatible with the observed event rate, is a viable model for the origin of GRBs.

### 0.6.2 Coincidences and other Astronomical Hints

Before examining the origin of GRBs, it is worthwhile to consider another astronomical phenomenon, SNRs, and see how one could have reached the right conclusions there. I have chosen SNRs since they seem to be the Newtonian cousins of GRBs. SNRs are observed as diffuse shells of optical and radio emission which originate from the interaction of supernovae ejecta with the ISM. Most SNRs are observed in a self-similar stage which is determined by two parameters: the energy of the ejected matter, which is $\approx 10^{51}$ ergs, and the density of the ISM. With so little information how do we know that these are really supernova remnants?

The chain of arguments is very simple. Supernovae observations show the ejection of several solar masses with velocities of tens of thousands km/sec. The corresponding kinetic energy is $\approx 10^{51}$ ergs - exactly in the right range. Additionally the birth rate of SNRs agree with the rate of supernovae!. Finally, there is a clear coincidence between pulsars that form in supernovae and SNRs.

Suppose now that we could not observe supernova directly and that we could not see the ejected material. Could we find that the observed diffuse shells result from core collapse without these observations? Surprisingly, the answer is yes. The amount of energy needed to produce SNR is quite large. This energy must have been released within a relatively short period of time. Only a few astronomical phenomena can do that - stellar core collapse that forms a neutron star is one. The discovery of pulsars has told us that neutron stars exist. The binding energy of a neutron star is larger than the kinetic energy required to produce an SNR, hence neutron star formation is a viable candidate for the source of SNRs. Estimates of the rate of pulsar formation and the birth rate of SNRs give a comparable rate. Thus, we have a phenomenon that can provide the energy (even though if we don't see supernovae we won't know that matter is ejected with the needed amount of energy) and it is taking place at a comparable rate. The final confirmation of the theory would come with the discovery of the Crab pulsar in the center of the remnant of the 1054 supernova. The existence of other pulsars in the centers of other SNRs will confirm that this was not a coincidence.

The situation with GRBs is remarkably similar to this conceptual toy model. At present we have very few clues on the process that causes GRBs. We know that the needed energy is $\approx 10^{50}$ ergs, which is rather close to the binding energy of a neutron star. This, in combination with the facts that the size of the source is quite likely less than $10^8$cm and that GRBs are catastrophic one-time events suggests that they are related to the formation of a compact object. The only other energetically feasible alternatives that I can see is the sudden release of the total rotational energy of a millisecond pulsar or the sudden release of the total magnetic energy of a neutron star with a $10^{15}$ Gauss magnetic field.

No mechanism has been suggested how to stop suddenly a rotating neutron star or a black hole. Thomson [57] suggested that GRBs are produced when a magnetic field of $10^{15}$ Gauss is suddenly destroyed. But there is no evidence that such magnetic fields exist in nature. It has been also suggested that GRBs occur in "failed" supernovae. However, it is not known that such events take place and if they do there is no idea what is their event rate (the lack of observations in Kamiokande puts an upper limit of less than once per ten years per galaxy - but this limit is very weak).

We are left with binary neutron star mergers ($NS^2Ms$) [21] or with a small variant a neutron star-black hole mergers [58]. These mergers take place because of the decay of the binary orbits due to gravitational radiation emission. The discovery of the famous binary pulsar PSR 1913+16 [59] have demonstrated unquestionably that this decay is actually taking place [60]. The discovery of other binary pulsars and in particular of PSR 1534+12 [61] have shown that PSR 1913+16 is not unique and that such systems are common. These observations suggest that $NS^2Ms$ take place at a rate of $\approx 10^{-6}$ events per year per galaxy [62, 63, 64], in amazing agreement with the GRB event rate [28, 65, 14]. Note that it has been suggested [66] that many neutron star binaries are born with very close orbits and hence with very short life time. If this idea is correct then the merger rate will be much higher. However, the short life time of those systems, which is the essence of this idea, makes it impossible to confirm or rule out this speculation.

$NS^2Ms$ result, most likely, in rotating black holes [67]. The process releases $\approx 5 \times 10^{53}$ ergs [68]. Most of this energy escapes as as neutrinos and gravitational radiation, but a small fraction of this energy suffices to power a GRB. The observed rate of $NS^2Ms$ is similar to the observed rate of GRBs. This is not a lot - but this is more than can be said, at present, about any other GRB model. It is also remarkably similar to our conceptual SNR toy model.

How can one proof or disproof this, or any other, GRB model? Theoretical studies concerning specific details of the model can, of course, make it more or less appealing. But in view of the fact that the observed radiation emerges from a distant region which is very far from the inner "engine" I doubt if this will ever be sufficient. Again, following our conceptual toy model, it seems that the only way to rule out or confirm any GRB model, will be via a coincident detection with another astronomical phenomenon, whose source could be identified with certainty. This brings us directly to another open question:
•*Is there a coincidence between GRB and any other phenomenon?*

NS$^2$Ms have two accompanying signals, a neutrino signal and a gravitational radiation signal. Both signals are extremely difficult to detect - but they provide a clear prediction of coincidence that could be proved or falsified sometime in the distant future.

## 0.7 WHY?

The last and probably most ambitious question is **why?** - that is why were GRBs put there in the sky? Put differently, what can GRBs tell us about the Universe that we live in? It is difficult to deal with this question when we don't know yet what is the origin of GRBs and we are not even certain where are they coming from. Still, it is worthwhile to speculate on the possible applications of GRBs to other branches of Astrophysics.

If, somewhat unexpectedly, it will turn out that GRBs are galactic, it will be the first indication that there is a population of stellar galactic objects that extend to distances of more than 100kpc. At present there is no other indication that there are objects at such distances. Furthermore, the distribution of these population must differ from the halo distribution inferred from dynamical studies of the Galaxy. This might have far reaching implications to theories of galactic structure. The origin of these population is an intriguing question that might teach us a lot about the galactic halo (if the sources are born in the halo) or about stellar processes in the galactic disk (if it will turn out that these objects are ejected from the disk).

If GRBs are cosmological they seems to be a relatively homogenous population of sources, with a narrow luminosity function (the peak luminosity of GRBs varies by less than factor of 10 [14, 69]) that are locates at relatively high $z$ values [28, 27, 70, 14]. Could GRBs be the holy grail of Cosmology and provide us with the standard candles needed to determine the cosmological parameters $H_0$, $\Omega$ and $\Lambda$? The answer is unfortunately no, at least not yet. Lacking any spectral feature, there is no indication what is the red shift of individual bursts and at present all that we have is the number vs. peak luminosity distribution. It turns out that this distribution is not sensitive enough to distinguish between different cosmological models (see Fig. 4) even when the sources are perfect standard candles with no source evolution [14]. Here, once more, we encounter the importance of finding counterparts to GRBs. Observation of such counterparts might provide us with additional parameter, such as the distance or the redshift, that when combined with the GRB data could determine the values of the cosmological parameters.

Finally I should mention an additional intriguing implications of the models that I have discussed so far. If GRBs are produces by ultra-relativistic particles it is possible that some particles escape and do not turn their energy to radiation and could be observed as cosmic rays[43]. With our recent understanding of energy conversion in shocks it was realized [71, 72, 73] that it is possible that the shocks that convert the kinetic energy to radiation also accelerate some of these particles to even higher energies [71, 72, 73]. Thus the events that produce GRBs might also generate cosmic rays. This is particularly intriguing as an explanation of the three mysterious $10^{20}$eV cosmic ray events [71, 72] discussed by Cronin [74] in this meeting.

## 0.8 CONCLUSIONS

It is not without reason that GRBs remained an unsolved problem for more than twenty years. The analysis that I have presented suggests that GRBs are the final outcome of a complicated process in which particles are first accelerated to ultra-relativistic energies and then convert their kinetic energy, via shocks, to the observed radiation. The fact that the observed radiation emerges from a region that is quite far from the internal engine that accelerate the particles and supplies the energy for the burst makes it quite difficult to find a conclusive test that will reveal the nature of this engine. It is clearly worthwhile to explore the nature of the conversion mechanism of kinetic energy to radiation, possible radiation mechanism and details of specific "engines" and acceleration mechanisms. However, I fear that the lack of any direct observation of the inner source region restricts our ability to proof or falsify most models. I view of this situation we should focus on location events that can produce the required energy and satisfy the temporal and size constraints and that are taking place at a comparable rate. Today, we have one such candidate, binary neutron star mergers. I believe that the search for coincident events in other wavelengths or other forms of emission should be the prime task of GRB research as this will be the best, clearest and most likely ultimate test of this and any other model.

I thank Ramesh Narayan, and Eli Waxman for many helpful discussions and Reem Sari and Ehud Cohen for allowing me to include unpublished results in this review. This research was supported by a BRF grant to the Hebrew University and by a NASA grant NAG5-1904.